\documentclass[twocolumn]{aastex631}

\submitjournal{AJ}

\shortauthors{Zasche et~al.} 

\graphicspath{{./}{figures/}}

\usepackage{tikz}
\usetikzlibrary{arrows,shapes,positioning,shadows,trees}

\tikzset{
  basic/.style  = {draw, text width=4cm, drop shadow, font=\sffamily, rectangle},
  root/.style   = {basic, rounded corners=2pt, thin, align=center, fill=blue!10},
  level 2/.style = {basic, rounded corners=6pt, thin,align=center, fill=pink!40, text width=11em},
  level 3/.style = {basic, thin, align=center, fill=white!60, text width=8em}
}

\usepackage[utf8]{inputenc}
\usepackage{dirtytalk}
\usepackage{natbib}

\begin{document}

\title{V907 Sco switched to the eclipsing mode again}

\correspondingauthor{Petr Zasche} \email{petr.zasche@email.cz}

\author[0000-0001-9383-7704]{Petr Zasche}
\author[0000-0002-6034-5452]{David Vokrouhlick\'y}
\affiliation{Astronomical Institute, Charles University, V Hole\v{s}ovi\v{c}k\'ach 2,
             CZ 18000, Prague 8, Czech Republic}
\author[0000-0002-8558-4353]{Brad N. Barlow}
\affiliation{Department of Physics and Astronomy, High Point University, One University Way,
             High Point, NC 27268, USA}
\author[0000-0002-0967-0006]{Martin Ma\v{s}ek}
\affiliation{Variable Star and Exoplanet Section, Czech Astronomical Society, Fri\v{c}ova 298,
             CZ 25165 Ond\v{r}ejov, Czech Republic}
\affiliation{FZU - Institute of Physics of the Czech Academy of Sciences, Na Slovance 1999/2,
CZ-182 21, Praha, Czech Republic}


\begin{abstract}
 V907~Scorpii is a unique triple system in which the inner binary component has
 been reported to switch on and off eclipses several times in modern history. In
 spite of its peculiarity, observational data on this system are surprisingly scarce.
 Here we make use of the recent TESS observations, as well
 as our own photometric and spectroscopic data, to expand the overall dataset and study the V907~Sco system in more detail.
 Our analysis provides both new and improved values for several of
 its fundamental parameters: (i) masses of the stars in the eclipsing binary are
 $2.74\pm 0.02$~M$_\odot$ and $2.56\pm 0.02$~M$_\odot$; and (ii) the third component
 is a solar-type star with mass $1.06^{+0.11}_{-0.10}$~M$_\odot$ (90\% C.L.), orbiting
 the binary on an elongated orbit with an eccentricity of $0.47\pm 0.02$ and a period of
 $142.01\pm 0.05$~days. The intermittent intervals of time when eclipses of the inner binary are
 switched on and off is caused by a mutual $26.2^{+2.6}_{-2.2}$ degree inclination of
 the inner- and outer-orbit planes, and a favourable inclination of about $71^\circ$
 of the total angular momentum of the system. The nodal precession period is $P_\nu=
 63.5^{+3.3}_{-2.6}$~yr. The inner binary will remain eclipsing for another $\simeq 26$~years,
 offering an opportunity to significantly improve parameters of the model.
 This is especially true during the next decade when the inner orbit inclination will
 increase to nearly $90^\circ$. Further spectroscopic observations are
 also desirable, as they can help to improve constraints on the
 system's orbital architecture and its physical parameters.
\end{abstract}


\section{Introduction} \label{intro}
Space-borne observations from the Kepler and Transiting Exoplanet Survey Satellite (TESS) missions
have assisted with the discover and study a particularly interesting class of compact hierarchical
triples (CHTs). These stellar systems are characterized by a small-enough ratio of the outer and
inner orbital periods. A plethora of dynamical effects at various timescales can be detected with
accurate photometric measurements for CHTs, and if properly interpreted with theoretical modeling,
they may bring interesting constraints on many physical parameters of the system. An excellent
review of this topic was recently published by \citet{bor2022}.

Few archetype CHT systems have been known even before the above-mentioned revolution. In these
cases, though, only the secular perturbations  related to nodal and pericenter precession were
typically detected. The short periods of these effects, sometimes even only a few decades, made
them belong to the CHT class \citep[see, e.g.,][for a review]{jetal2018}.

V907~Sco is a particularly interesting member of this class of objects. While occasionally
observed in the second half of the 20th century, and first cryptically reported to be an eclipsing
binary by \citet{sko1964}, a truly comprehensive study of V907~Sco was presented by
\citet{letal1999}. We encourage the interested reader to learn about the sometimes tangled history
of the V907~Sco analysis from the accumulated photometric data in this reference (principally
Sec.~1). Here we only mention that \citet{letal1999} were the first to fully understand the
background story of V907~Sco by proposing that it is in fact a triple in which the previously
known binary is accompanied by a third star in a sufficiently close orbit. This was necessary to
explain the surprisingly short changing periods during which the binary ceased to be eclipsing.
Unfortunately \citet{letal1999} could not continue to collect accurate photometry of the eclipsing
component, since it was not eclipsing during the period of their work. Their important
contribution to the observational evidence about V907~Sco thus consisted in taking the first
series of the spectroscopic data. While useful in several aspects (e.g., confirming precisely the
period of the eclipsing binary, determining the semiamplitudes of the radial velocity curves, and
noting the narrowness of the spectral lines), \citet{letal1999} were unlucky by having available
only a sparse set of spectra spread over an interval of 11 years. Their analysis was therefore
ill-suited for such a fast-evolving system and led to a few mistakes. First, they attempted to
determine masses of the stars in the inner binary but chose an incorrect orbital inclination at
the reference epoch of their observations. As a result, their reported masses were slightly
incorrect (Sec.~\ref{obs}). More importantly though, the sparsity of their data led them to
determine basic parameters of the outer orbit, the period and orbital eccentricity, incorrectly
(see again Sec.~\ref{obs}). To their credit, however, we finally note that \citet{letal1999} also
considered the possibility that V907~Sco belongs to the open cluster M7. Comparing the estimated
distance to both M7 and V907~Sco, though the latter arguably rather approximate at that moment,
they rightly rejected the association. Their conclusion was recently confirmed with astrometric
measurements from Gaia DR3 showing that  V907~Sco has a parallax of $2.279\pm 0.033$~mas
\citep{2022arXiv220800211G}. This translates to a distance of $439\pm 6$~pc, well beyond that of
M7.

The rather rough and partially incorrect analysis of \citet{letal1999} let these authors to
predict that the eclipses of V907~Sco will start again in $2030\pm 5$ in their concluding
Section~4. However, \citet{bor2022} noted that the recent observations of TESS have already
revealed an onset of V907~Sco eclipses in the Sector~13 data (in mid 2019), and confirmed their
expected trend towards deeper eclipses in the Sector~39 data (in mid 2021). Clearly, the situation
demands a new analysis of V907~Sco, since the recent emergence of its eclipsing state alone proves
that some assumptions and/or data treatment in \citet{letal1999} have not been correct. This is
the primary purpose of this work. However, realizing that past data (especially the pre-TESS
photometry) could be improved upon, we also aim to extend the empirical data set on this unusual
triple system with new observations.

The paper is organized as follows. In Sec.~\ref{obs} we describe photometric and spectroscopic
observations available to us, including archival, previously published, and new data we obtained
in the 2022 season. Realizing that several orbital parameters are evolving quickly, we split the
analysis into two steps. First, still in Sec.~\ref{obs}, we focus on the bulk of the 2021-2022
data. They conveniently cover a period slightly longer than the outer orbit period, allowing us to
properly characterize it, but short enough that the changes of the orbital parameters are minimal.
We can thus use plain Keplerian orbits for the data analysis here. This allows us to determine
orbital and physical parameters of the inner binary in the V907~Sco system (included the masses
and radii of the stars). With this information available, we can thus use the archival photometry
and derive inclination changes of the eclipsing binary over more than a century. In Sec.~\ref{sec}
we develop a model accounting for mutual gravitational interactions of all three stars in the
system to describe the observed variations of the orbital elements over long-term period (many
cycles of the outer orbit). Confrontation with the data allows us to determine some parameters of
the third star (including its mass and tilt of its orbital plane with respect to the orbital plane
of the inner binary). Finally, in Sec.~\ref{concl}, we outline future prospects to study V907~Sco
in even more detail.

\section{Observations and short-term analysis}   \label{obs}
Both photometric and spectroscopic observations are available for V907~Sco. The photometry falls
into the following three categories: (i) archival data sparsely spanning more than a century, some
of which have also been mentioned by \citet{letal1999}, (ii) the above-mentioned TESS
observations, and (iii) our new data from the 2022 season. As for the spectroscopy, we conducted
an intensive observing campaign in 2022, knowing how quickly the V907~Sco system evolves. Such
data have the advantage of minimizing long-term evolutionary effects in spectra. There are also
radial velocity measurements published by \citet{letal1999} (their Table~3). However, they are
spread over a time interval of nearly $11$ years, and thus their analysis strongly depends on the
ability to consistently include the evolutionary model of the whole system. For that reason, we do
not consider them here, but save their analysis for the future work.
\begin{figure}
  \centering
  \includegraphics[width=0.47\textwidth]{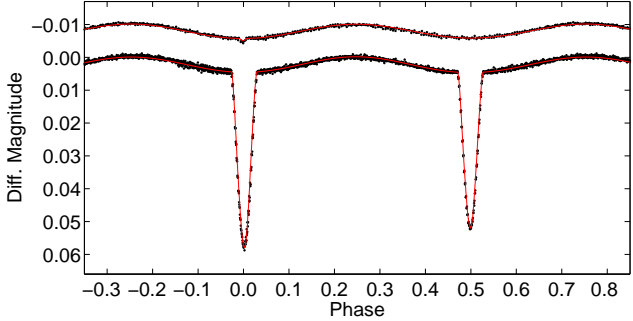}
  \caption{TESS photometry of V907~Sco from Sectors 39 (lower plot) and 13 (upper plot), each spanning about $27$ days and folded on the $3.7762049$~day period of the eclipsing binary.
  The ordinate is differential magnitude, which has been shifted by $0.01$~mag for the Sector~13 data for better visualization.
  The red lines are the respective best-fits obtained using the {\sc PHOEBE} code (Table~\ref{LCaRVparam}). In the Sector 13 data, only a very subtle dip is noticed at the zero phase of the primary eclipse, implying the stellar configuration seen by the observer was nearly grazing.}
  \label{LC_V907Sco}
\end{figure}

\begin{figure}
 \includegraphics[width=0.47\textwidth]{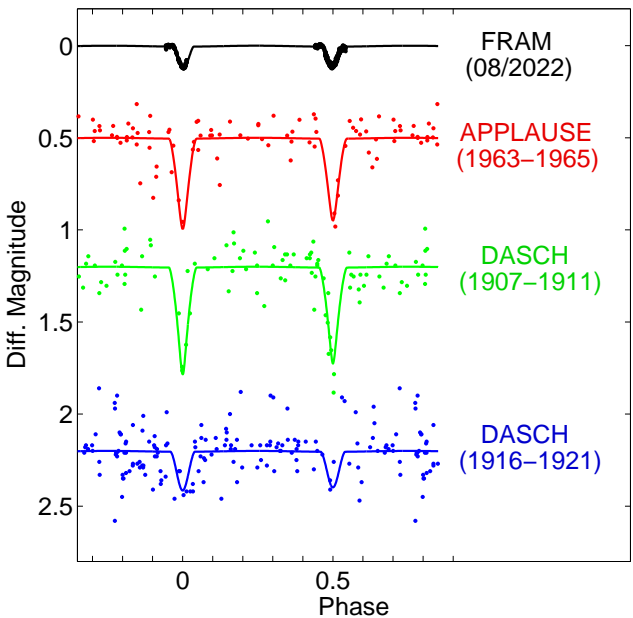}
 \caption{Sample light curves from different epochs: symbols are data, solid line is the best-fit model (the corresponding inclination values $i_1$ are listed in Table~\ref{inclination}). The abscissa is the phase of $P_1$ cycle, and the ordinate is the differential magnitude, with a shift for sake of visibility. Top data are from modern instrumentation (FRAM observations in August 2022), bottom three examples are from various archival sources.}
 \label{LCs}
\end{figure}

\subsection{Photometric observations}\label{phot}
In what follows, we describe each photometric data set, in order of decreasing precision.
\smallskip

\noindent{\em TESS photometry.-- }We searched all TESS sectors for field-of-views containing
V907~Sco and found that the object was observed twice, in Sectors 13 and 39 \citep[V907~Sco =
TIC~261030551; see also][]{bor2022}. Each of them represents about 27~day long period of
continuous data, with an approximately day-long data downlink gap in the middle. Target photometry
was extracted using the publicly available {\sc lightkurve} package
\citep[e.g.,][]{2018ascl.soft12013L}. Coincidentally, and of high value to this study, Sector~13
observations began at the onset of the eclipses, as the light curve shows only a very shallow
signal/dip at the primary eclipse phase. The exposure cadence was about 30 minutes. The Sector~39
data are superior in their quality and value for our analysis. Not only do they have a more rapid
cadence (10 minutes), but they were taken at the moment when the stars in the inner binary of
V907~Sco exhibited sufficiently deep eclipses (about 0.06 mag in the case of the primary eclipse)
for a thorough analysis. With the expected millimagnitude-level accuracy of the TESS photometry
\citep[e.g.,][]{2015JATIS...1a4003R}, the Sector~39 data are by far the most precise photometry we
possess on the V907~Sco system. Consequently, they provide important constraints on fundamental
orbital and physical parameters of the inner binary. Figure~\ref{LC_V907Sco} shows the TESS light
curves from both sectors, phase-folded on the best-fitting mean period $P_1$ of the binary. We
note that our definition of the zero phase is shifted by half an orbital period compared to the
definition of \citet{letal1999}. This is because we prefer to associate the zero phase with the
deeper eclipse, when the smaller (less massive) star in the binary eclipses the larger (heavier)
star.

We used the well-tested software package {\sc PHOEBE} \citep{2005ApJ...628..426P},  based on the
\cite{1971ApJ...166..605W} code, to obtain fundamental parameters of the inner binary system using
(i) TESS Sector~39 photometric data, and (ii) our new spectroscopic data described in
Sec.~\ref{spectra}. Ideally, both datasets should have been taken at the same epoch, but they were
not. Luckily, the one year difference between them is short. Figure~\ref{fig2} shows that the
binary inclination $i_1$ changed by only $\simeq 1^\circ$ from 2021 to 2022, and this makes the
semiamplitudes of the radial velocities change by about $0.3$\%. This is a negligible change in
our solution. Our modeling required only minimum assumptions, such as rotational synchronicity of
the stars in the binary ($F_i=1$), while other parameters were left adjusted. This concerns not
only the primary ones, such as stellar masses and radii, but also more subtle ones, such as the
gravity brightening or albedo. The latter primarily influences the ellipsoidal modulation of the
photometric signal outside the eclipses (Fig.~\ref{LC_V907Sco}). The results are summarized in
Table~\ref{LCaRVparam}. The masses, $2.56$~M$_\odot$ and $2.74$~M$_\odot$, are about $8$\% larger
than previously suggested by \citet{letal1999}. This is because these authors assumed an
inaccurate value of the inclination $i_1$ of the binary orbital plane at the effective epoch of
their spectroscopic observations. The corresponding radii are $2.53$~R$_\odot$ and
$2.68$~R$_\odot$, consistent with the values expected for the B9V and B8V main sequence stars. The
contribution of the third light in the TESS passband is about $6$\%, and the binary has a small,
but well detectable, orbital eccentricity of about $0.008$. Both values are found to be consistent
with our modeling discussed below (see Sec.~\ref{concl}).

\begin{deluxetable}{lc}
\tablenum{1} \tablecaption{Orbital and physical parameters of the  inner binary in V907~Sco as
derived from the combined lightcurve and radial velocity fitting. As for the orbital inclination
values, see Table~\ref{inclination}. The reference epoch $JD_0$ of the solution corresponds to the
center of the first primary eclipse in the TESS Sector~39 data. \label{LCaRVparam}}
\tablewidth{0pt} \tablehead{
  \colhead{Parameter} & \colhead{Value}
}
\startdata \rule{0pt}{3ex}
    $\,\hspace{15mm}\,$Orbital & parameters :$\,\hspace{15mm}\,$   \\
  $P_1$ [d] & 3.7762049 $\pm$ 0.0000122 \\
  $a_1$ [$R_\odot$]      & 17.79 $\pm$ \,0.05         \\
  $e_1$                  & 0.0080 $\pm$ \,0.0025      \\
  $\omega_1$ [deg]       &  122.6 $\pm$ \,3.3         \\
  $v_\gamma$ [km s$^{-1}$]      & -17.33 $\pm$ \,0.24        \\
  $JD_0$ [$JD-2450000$]  & 9362.3272 $\pm$ \,0.0023   \\
 \hline
    $\,\hspace{15mm}\,$Physical & parameters :$\,\hspace{15mm}\,$   \\
 \rule{0pt}{3ex}
 $q\, (=\frac{m_{1b}}{m_{1a}}$) & 1.072 $\pm$ \,0.007 \\
  $m_{1a}$ [$M_\odot$]      & 2.56 $\pm$ \,0.02  \\
  $m_{1b}$ [$M_\odot$]      & 2.74 $\pm$ \,0.02  \\
  $R_{1a}$ [$R_\odot$]      & 2.53 $\pm$ \,0.04  \\
  $R_{1b}$ [$R_\odot$]      & 2.68 $\pm$ \,0.04  \\
  $T_{1a}$ [K]              &  10004 $\pm$ \,35  \\
  $T_{1b}$ [K]              &  10000 (fixed)     \\
  $M_{\mathrm{bol},1a}$ [mag] & 0.324 $\pm$ \,0.012 \\
  $M_{\mathrm{bol},1b}$ [mag] & 0.204 $\pm$ \,0.011 \\
  $\log g_{1a}$ [cgs]       & 4.039 $\pm$ \,0.010  \\
  $\log g_{1b}$ [cgs]       & 4.021 $\pm$ \,0.012  \\
  $L_{1a}$ [\%]             & 44.8 $\pm$ \,0.6     \\
  $L_{1b}$ [\%]             & 49.3 $\pm$ \,0.7     \\
  $L_2$ [\%]                & 5.9 $\pm$ \,1.4      \\ \hline
  \enddata
\end{deluxetable}
\smallskip

\noindent{\em FRAM photometry.-- }The most recent data, taken between April and September 2022,
had the purpose of confirming the trend of the eclipse depth increase (and thus the inclination of
the binary orbital plane). They were obtained with the F/(Ph)otometric Robotic Atmospheric Monitor
(FRAM) 30~cm telescope at Pierre Auger Observatory, Argentina
\citep[e.g.,][]{PA_Collaboration_2021}. This is an Orion ODK 300/2040mm system equipped with the
CCD camera MII G4-16000; standard reduction routines using dark frames and flat fields were used.
The small aperture of the instrument results in the FRAM photometry having less precision than the
TESS photometry; a typical uncertainty in an individual frame is $\simeq 0.01$ magnitude. However,
given the fact that the current eclipse depth is already about ten times larger, the FRAM data are
still very useful (see the top curve on Fig.~\ref{LCs} and Fig.~\ref{fig_depth}). We obtained
observations at four epochs in 2022, clearly demonstrating the expected progression towards deeper
eclipses. Importantly, when this trend is mapped to the increasing inclination of the binary
orbital plane (Table~\ref{inclination}), the time evolution is not linear. Indeed, in
Sec.~\ref{res_sec} we find reasons for this non-linearity in the dynamical model.

\begin{deluxetable*}{ccccc}
\tablenum{2} \tablecaption{Orbital inclination values $i_1$, and their uncertainty
  $\Delta i_1$, determined from different datasets (last column). The fourth
  column indicates the interval of time during which data were collected (in heliocentric Julian date, $2400000$ subtracted), the first column is the reference epoch of the inclination used in Sec.~\ref{res_sec}.
\label{inclination}} \tablewidth{0pt} \tablehead{
  \colhead{mid HJD} & \colhead{$i_1$ (deg)} &
  \colhead{$\Delta i_1$ (deg)} &
  \colhead{HJD range} & \colhead{Source}
}
\startdata
  \rule{0pt}{3ex}
  & \multicolumn{3}{c}{ -- Archival data --} & \\
  \rule{0pt}{3ex}
  \hspace*{-5pt}
  15207.0  &   86.0     &  7.3   &   14018 -- 16396   & DASCH  \\
  16835.5  &   86.6     &  2.9   &   16178 -- 17494   & DASCH  \\
  18312.0  &   88.0     &  1.8   &   17685 -- 18939   & DASCH  \\
  19787.5  &   87.3     &  1.6   &   19149 -- 20426   & DASCH  \\
  20910.0  &   83.7     &  3.4   &   20304 -- 21516   & DASCH  \\
  21793.0  &   81.2     &  3.7   &   21006 -- 22580   & DASCH  \\
  38420.5  &   86.23    &  1.85  &   38196 -- 38645   & APPLAUSE  \\
  \rule{0pt}{3ex}
  & \multicolumn{3}{c}{ -- Modern data --} & \\
  \rule{0pt}{3ex}
  \hspace*{-5pt}
  58675.0  &   73.37    &  0.25  &   58670 -- 58680   & TESS Sector 13  \\
  59365.3  &   76.57    &  0.02  &   59362 -- 59369   & TESS Sector 39a \\
  59383.9  &   76.61    &  0.02  &   59380 -- 59387   & TESS Sector 39b \\
  59673.8  &   77.78    &  0.10  &   59673 -- 59674   & FRAM  \\
  59771.0  &   78.11    &  0.08  &   59762 -- 59780   & FRAM  \\
  59794.7  &   78.45    &  0.04  &   59794 -- 59795   & FRAM  \\
  59812.0  &   78.55    &  0.04  &   59811 -- 59813   & FRAM  \\
\enddata
\end{deluxetable*}
\smallskip

\noindent{\em Archival photometry.-- }In addition to our recent photometric data, we also utilized
archival resources. Obviously, this is motivated by the reported intermissions in V907~Sco
eclipses \citep{letal1999}, and thus we seek the ability to describe them. We used data from: (i)
the Archives of Photographic PLates for Astronomical USE (APPLAUSE) project of digitalizing the
old photographic plates available online\footnote{\url{https://www.plate-archive.org/query/}}
\citep{APPLAUSE}, and (ii) the Digital Access to a Sky Century at Harvard (DASCH) database of
photographic plates from Harvard observatory\footnote{\url{http://dasch.rc.fas.harvard.edu/}}
\citep{2013PASP..125..857T}. Given the fact that DASCH plates cover an interval of several
decades, we could even parse them into several independent intervals in between 1900 and 1920. In
spite of their rather poor quality (see the bottom curves on Fig.~\ref{LCs}), these data are still
very useful as they approximately tell us trends in the inclination variations over a long
time-span. Obviously, we downscale the weight of these data by assigning large uncertainty to the
derived inclination values (Table~\ref{inclination}).

Despite our efforts, we unfortunately did not find more photometric datasets from existing survey
programs that would provide useful information about V907~Sco. An example is ASAS-sn in 2022, in
which observations in the V907~Sco field are subject to a spurious instrumental scatter of about
1.5 mag.
\begin{figure}
 \includegraphics[width=0.46\textwidth]{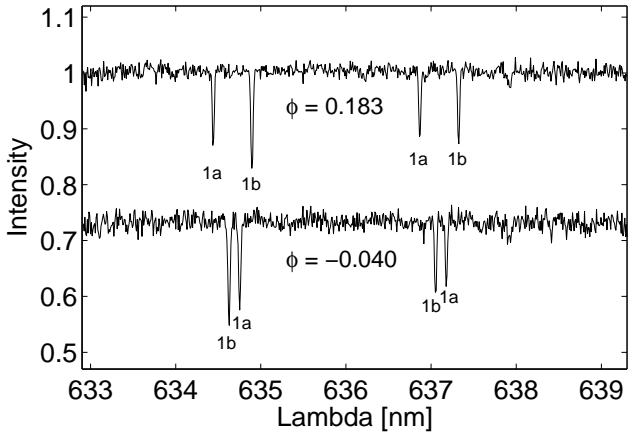}
 \caption{An example of two spectra taken at different phases $\phi$ of the inner binary motion (vertically shifted for visibility) obtained by CHIRON spectrograph. Labels 1a and 1b identify the stellar components in the binary. A narrow window in the red part of the spectrum around the SiII lines (634.7 and 637.1 nm) is shown.}
 \label{Spectra}
\end{figure}
\smallskip

\noindent{\em Compilation of the inclination values at different epochs.-- } Having calibrated the
physical parameters of stars in the inner binary from the TESS photometry, we can now fix them and
use other photometric data to determine only the inclination of the eclipsing binary orbital plane
at their epoch. For this task we used the {\sc PHOEBE} code as above. Note also that ephemerides
for the eclipses are determined accurately enough for our purpose: a fractional uncertainty of
$\Delta P_1/P_1\simeq 3\times 10^{-6}$ expands to a $\simeq 0.03$ phase uncertainty after a
century-long interval (i.e., roughly representing $10^4$ cycles of the binary). This is due to the
extreme accuracy of the TESS photometry. Nevertheless, we adjusted the epoch of the primary
minimum individually in each of the datasets, such that the fit to the model is optimized. As
expected, this adjustment turned out to be minimal.

The results are summarized in Table~\ref{inclination} for both the less accurate archival data and
the much more precise modern data. From the TESS Sector~13 the value was only roughly estimated,
while for the TESS Sector~39 the data were split evenly into two parts (roughly two weeks each),
and the inclination was determined for each independently. The two $i_1$ values are indeed
statistically different and match well the expected trend (Sec.~\ref{res_sec}). We are aware of
the fact that the TESS data suffer from light contamination from nearby sources and its level may
change between different sectors, or even during one sector of data. Due to strong correlation
between the inclination and additional stray light this may play some role. However, the
difference would be only very small, and the principal results of the analysis when using one
single averaged point or two instead would remain almost unchanged. All inclination values are
later used in Sec.~\ref{sec} to constrain parameters of the third component in the system from its
gravitational perturbations of the inner binary.

\begin{deluxetable}{lc}
\tablenum{3} \tablecaption{Parameters of the outer orbit in V907~Sco from available radial
velocity data in 2022. The semi-amplitude $K$ corresponds to the motion of the inner binary center
of mass with respect to the center of mass of the whole system.
 \label{RVpar3}}
\tablewidth{0pt} \tablehead{
  \colhead{Parameter} & \colhead{Value}
}
\startdata
 $P_2$ [d]        & 142.01 $\pm$ \,0.05  \\
 $e_2$            & 0.467 $\pm$ \,0.019  \\
 $\omega_2$ [deg] & 201.4 $\pm$ \,2.7    \\
 $K$ [km s$^{-1}$]  & 13.75 $\pm$ \,0.82   \\
\enddata
\end{deluxetable}

\begin{figure}
 \includegraphics[width=0.47\textwidth]{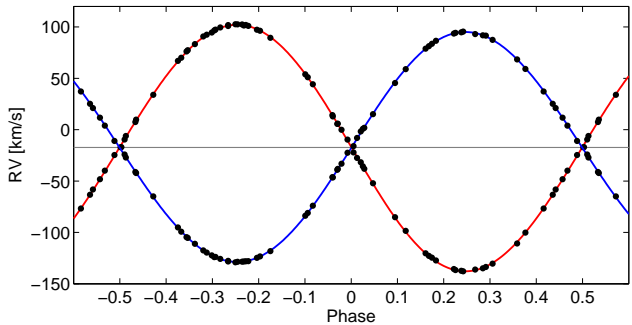}
 \caption{Disentangled radial velocities at the period $P_1$ for the components in the eclipsing binary using CHIRON spectra from 2022. Black symbols are the individual data, red/blue lines are the Keplerian best-fit for the primary/secondary components with parameters listed in Table~\ref{LCaRVparam}. The abscissa is the phase of the binary cycle (with a zero value at the epoch of the primary eclipse). The grey horizontal line is the systemic velocity of $v_\gamma\simeq -17.3$ km~s$^{-1}$.}
 \label{RV_V907Sco}
\end{figure}
\begin{figure}
  \centering
  \includegraphics[width=0.47\textwidth]{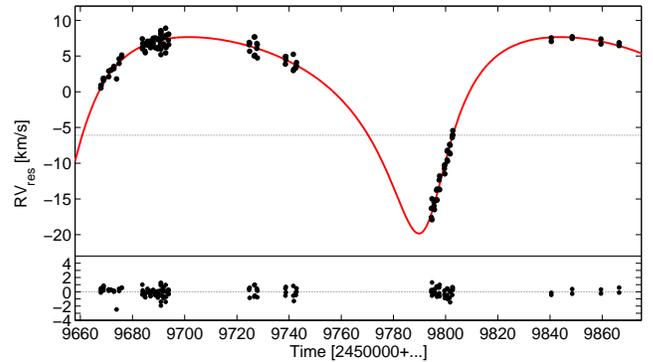}
  \caption{Upper part: disentangled radial velocities at the period $P_2$ for the center of mass of the inner binary. The individual data points shown with black symbols. The red line is the Keplerian best-fit with parameters listed in Table~\ref{RVpar3}; the grey horizontal line is at $K\,e_2\cos\omega_2\simeq -6$ km~s$^{-1}$, a center to which semi-amplitude $K$ is referred to. The abscissa is the Julian date with 2450000 subtracted. Bottom part: residuals of the radial velocities after motion of the inner and outer orbit has been subtracted from the data.}
  \label{RV3plot}
\end{figure}

\subsection{Spectroscopic observations} \label{spectra}
Our new spectroscopic observations of V907~Sco were obtained using the CHIRON echelle
high-resolution spectrograph ($\mathrm{R}\simeq 28000$) mounted on a $1.5$~m SMARTS telescope at
CTIO observatory located on Cerro Tololo, Chile  \citep[e.g.,][]{2013PASP..125.1336T}. Altogether,
we have $66$ spectra taken in between March and October 2022. Typical exposure times were $480$~s,
and the usable spectral range covered a wavelength range of $450$ to $890$~nm.

The whole reduction process including rectification of the spectra, identifying the components,
and measuring their radial velocities, all of which were accomplished using the {\sc reSPEFO 2}
package\footnote{\url{https://astro.troja.mff.cuni.cz/projects/respefo/}}
\citep{2020A&A...639A..32H}. The resulting values at each epoch observed are listed in
supplementary materials (Table~A1). For most of the spectra, we were able to identify a pair of
lines that belong to the two stars in the inner binary of V907~Sco (corresponding mainly to Fe and
Si; see Fig.~\ref{Spectra} for two examples at different phases of the binary motion).
Unfortunately, we could not identify spectral lines of the third star in the system \citep[see
also][]{letal1999}.  This non-detection can be explained by a low luminosity (and thus mass)
compared to the stars in the inner binary. We estimate a $\simeq 5$~magnitude difference between
the third star and the binary, or a few percent contribution to the total luminosity of the system
at maximum. Given the masses derived in Table~\ref{LCaRVparam} for the components in the binary,
we estimate an upper mass limit on the third star of $\simeq 1.05$~M$_\odot$. This nicely matches
the results in Sec.~\ref{res_sec}, where we infer the third component mass from its dynamical
effects on the binary.

As mentioned in the previous Section, we used the code {\sc PHOEBE} to analyse simultaneously the
TESS-Sector~39 photometry and our derived radial velocities in order to solve for the orbital and
physical parameters of the eclipsing binary (i.e., all signal at the $P_1$ period). This resulted
in the values presented in Table~\ref{LCaRVparam} and the velocity curve fits shown in
Fig.~\ref{RV_V907Sco}. The data are matched by the model quite well. Next we analysed the residual
signal in the measured radial velocities and expected to find the $\simeq 99$ day period of the
outer orbit reported by \citet{letal1999}. To our surprise, we obtained something else (see
Fig.~\ref{RV3plot}). When fitting a Keplerian orbit to the velocity residuals, we find a long
period of $P_2\simeq 142$~days with a significant eccentricity of $e_2\simeq 0.47$.
Table~\ref{RVpar3} presents the resulting parameters. The model fit is rather good, with the final
residuals having a dispersion of only $0.64$ km~s$^{-1}$ (at the level of the radial velocity
measurement uncertainty) and no further signals being detectable (see Supplementary materials,
Table~A1). We unfortunately lack data near the minimum and maximum of the radial velocity curve
for the outer orbit (this happened partly due to bad weather and partly due to overload of the
telescope by other projects). This leaves us with parameter uncertainties in Table~\ref{RVpar3}
slightly larger than we would wished. As an example, the minimum mass of the third component, as
derived from the mass function, ranges between $0.95$~M$_\odot$ and $1.09$~M$_\odot$. We only note
that this value is satisfactorily compatible both (i) with the maximum third component mass from
the absence of its spectroscopic evidence and the third light contribution, and (ii) also with the
solution based on photometric data in Sec.~\ref{res_sec}.

Finally, we note that the large value of the eccentricity of the outer orbit is still comfortably
within stability limits. Assuming stellar masses in the eclipsing binary derived above, and
$m_2\simeq 1$~M$_\odot$ (Sec.~\ref{res_sec}), we find that the system is stable up to $e_2^{\rm
stab}\simeq 0.63$ according to the criteria in \citet{ek1995} and \citet{ma2001}.

\section{Long-term evolution of the system} \label{sec}
In this section, we analyse the observational results of the V907~Sco system collected above and
summarized in Table~\ref{inclination} using a simple, but adequately accurate model. This allows
us to describe changes in the system's architecture and its parameters, such as orbital
inclinations of both inner and outer orbits, over a timescale of decades. Before discussing the
results, we review the basic aspects of our theoretical approach and relevant variables.
Additional details can be found in \citet{hs2022} or \citet{jetal2018}, while an interested reader
may also want to revisit a classical series of papers by \citet{s1975,s1982,s1984}. We find it
useful to start with a simple analytic secular model in Sec.~\ref{theory1}. Effects with long
period $P_2$ are not described by this part, but its strength stems from analyticity, which
effectively helps us pre-constrain the model parameters. Next, in Sec.~\ref{theory2}, we also
include the long-period effects using a computationally more demanding numerical model.

\subsection{Simple analytical model for the secular part} \label{theory1}
We first introduce notation and important parameters used throughout this section: (i) $m_{1a}$
and $m_{1b}$ are masses of the two stars in the eclipsing binary (inner orbit), defining their
total and reduced masses simply as $M_1 = m_{1a} + m_{1b}$ and $\mu_1 = m_{1a} m_{1b}/M_1$, (ii)
$m_2$ is the mass of the third star in the system (constituting the outer orbit with respect to
the barycenter of the inner orbit), and the total mass is $M_2=M_1+m_2$. We denote also the mean
orbital periods $P_1$ and $P_2$ of the inner and outer orbits, or alternatively also the
corresponding mean motions $n_1=2\pi/P_1$ and $n_2= 2\pi/P_2$. Finally, we assume the inner orbit
is circular ($e_1=0$), and the outer orbit may have an arbitrary eccentricity $e_2$, defining also
a convenient parameter $\eta_2=\sqrt{1-e_2^2}$.

The principal variables of interest are the orbital angular momenta ${\bf L}_1$ and ${\bf L}_2$ of
the inner and outer orbits, which secularly change due to gravitational interactions in the
system. When the latter is restricted to the quadrupole coupling, justifiable in the V907~Sco
system for which the mass ratio $q$ of the stars in the inner system has been found close to
unity, (i) the angular momentum magnitudes $L_1$ and $L_2$ are secularly conserved, and (ii) the
unit vectors ${\bf l}_1$ and ${\bf l}_2$, defining orientation of the respective angular momenta
in space, steadily precess about the conserved total angular momentum of the triple ${\bf L} = L\,
{\bf l} = {\bf L}_1+{\bf L}_2$ (here we neglect small contributions of the rotational angular
momenta of the three stars, again justifiable for the V907~Sco system, where the narrow spectral
lines of the components in the eclipsing system imply their slow rotation). Additionally, as the
angular momentum vectors of the inner and outer system roll on the respective cones about ${\bf
L}$, their mutual configuration is fixed such that their mutual angle $J$ defined by $\cos J =
{\bf l}_1\cdot {\bf l}_2$ (also depicting the angle between the orbital planes of the inner and
outer orbits), remains secularly constant.

It should be recalled that within the point-mass, three-body problem adopted here, the simple
solution outlined above holds when $J\leq 40^\circ$ or $J\geq 140^\circ$. In between these values
a more complicated, Kozai-Lidov regime occurs \citep[e.g.,][]{s1982,fl2010}. Luckily, this
complication is of no concern for V907~Sco. First, we show below that the observations require
configuration for which $J$ is restricted to a narrow range of values about $\simeq 26.2^\circ$
(Sec.~\ref{res_sec}). Second, the long-term stable triples often rearrange such that the
Kozai-Lidov regime is effectively eliminated by mutual (tidal) interaction of the stars in the
inner orbit \citep[e.g.,][]{s1984}; this even allows systems with $J= 90^\circ$ to exist, the
Algol case representing the archetype member of this class \citep[e.g.,][]{baron2012}.

Mathematical representation of the steady precession state of the inner and outer angular momenta
reads simply
\begin{equation}
 {\bf l}_1(t)  =  {\bf {\hat l}}_1\cos\alpha + \left({\bf l}\times {\bf {\hat l}}_1\right)
  \sin\alpha
   + {\bf l} \left({\bf l}\cdot {\bf {\hat l}}_1\right) \left(1-\cos\alpha
  \right)\; , \label{i1}
\end{equation}
and similarly for ${\bf l}_2(t)$ (recall ${\bf l}$ is the unit vector of the conserved total
angular momentum of the system). Here, ${\bf {\hat l}}_1$ is the unit vector of the inner orbit
angular momentum at some arbitrary reference time $T_0$ (i.e., ${\bf {\hat l}}_1 = {\bf
l}_1(T_0)$), and $\alpha = \nu (t-T_0)$, where $\nu$ is the fundamental precession frequency. Some
authors are satisfied with $\nu$ as a free, solved-for parameter, but the analytical theory offers
an important step further, namely to interpret $\nu$ in terms of more fundamental parameters such
as $J$ and masses of stars in the triple system. Remaining still on the grounds of the quadrupole
coupling approximation, we have \citep[e.g.,][]{s1975,bv2015}
\begin{equation}
 \frac{\nu}{n_2} = -\frac{3}{4\eta_2^3} \frac{m_2}{m_2+M_1} \frac{n_2}{n_1}
  \cos J \sqrt{1+\gamma^2 + 2\gamma\cos J}\; , \label{prec}
\end{equation}
where
\begin{equation}
 \gamma = \frac{\mu_1}{m_2 \eta_2}\left( \frac{m_2+M_1}{M_1} \frac{n_2}{n_1}
  \right)^{1/3} \label{gama}
\end{equation}
is the ratio of the inner and outer orbit angular momenta (i.e., $\gamma= L_1/L_2$). Typical
triple systems have $\gamma\ll 1$, as the outer orbit carries much larger share of the system's
angular momentum. However, V907~Sco belongs to a small group of outlier cases in this respect.
This is because (i) it is rather compact with $P_2/P_1 \simeq 37.6$, and (ii) the third component
is most likely a solar-type star (Sec.~\ref{res_sec}). Adopting the canonical value $M_1\simeq
5.3$~M$_\odot$, the plausible $m_2\simeq 1$~M$_\odot$ mass implies $\gamma\simeq 0.47$, thus a
partial share of the inner and outer orbits on the total angular momentum. Large variations of the
inner orbit inclination $i_1$ (witnessed by the intermittent periods of eclipses and no-eclipses)
must therefore be accompanied with certain variations of the inclination $i_2$ of the outer orbit.
The analytical solution (\ref{i1}) provides this result readily: orienting the reference system as
usual, such that the $XY$ plane coincides with the sky-plane, the inclination $i_1(t)$ of the
eclipsing system is simply $\cos i_1(t) = {\bf l}_1(t)\cdot {\bf e}_3$, where ${\bf e}_3^{\rm T} =
(0,0,1)$ (and similarly for the outer orbit, with just replacing ${\bf l}_1(t)$ by ${\bf
l}_2(t)$).

It is useful at this moment to overview free parameters of the analytical model whose values are
to be adjusted (solved for) when confronting with the data (Sec.~\ref{res_sec}). Choosing the
reference epoch $T_0$, we first have ${\bf {\hat l}}_1$ and ${\bf {\hat l}}_2$ (normal unit
vectors to the planes of the inner and outer orbits) to select (note their composition with
magnitudes $L_1$ and $L_2$ of the angular momenta define the unit vector ${\bf l}$ of the total
angular momentum of the system). Being unit vectors, each of the two would apparently depend on
two solved-for parameters. But given the nature of the data we have, photometric and spectroscopic
sets, there is an arbitrariness in selection of the reference direction in the sky-plane. As a
result, there are only three solved-for parameters needed to characterize the unit vectors ${\bf
{\hat l}}_1$ and ${\bf {\hat l}}_2$. For instance, we may assume
\begin{equation}
 {\bf {\hat l}}_1 = \left(\begin{array}{c}
                      0 \\  -\sin {\hat {\i}}_1 \\ \phantom{-}\cos {\hat {\i}}_1 \\
                      \end{array}\right) \; ,
\end{equation}
where ${\hat {\i}}_1$ is the inner orbit inclination at $T_0$, and
\begin{equation}
 {\bf {\hat l}}_2(0) = \left(\begin{array}{c}
                        \phantom{-} \sin {\hat {\i}}_2 \sin{\hat \Omega}_2 \\
                        -\sin {\hat {\i}}_2 \cos{\hat \Omega}_2 \\
                        \cos {\hat {\i}}_2 \\
                      \end{array}\right) \; ,
\end{equation}
where ${\hat {\i}}_2$ is the outer orbit inclination at $T_0$, and ${\hat \Omega}_2$ is the nodal
phase tilt about ${\bf e}_3$ of the outer orbit with respect to the inner orbit. Therefore, we
have three solved-for model parameters $({\hat {\i}}_1, {\hat {\i}}_2,{\hat \Omega}_2)$ of the
geometrical nature. If only photometric data were available, one could complement this set of
three parameters with the precession frequency $\nu$ to be the fourth solved-for parameter, and
use the inner system inclination changes to determine their values \citep[a formulation dating
back to][]{s1975}. A partial step forward was introduced by \citet{jetal2018}, who used the
photometrically-derived constraint on $\nu$, and formula Eq.~(\ref{prec}) to set constraints on
further physical parameters, such as $m_2$ and often unknown $P_2$ for their systems.

Here, though, we are in a slightly more comfortable situation with both photometric and
spectroscopic data available, and we can thus afford more general formulation (though a really
ambitious approach needs to wait when more accurate data are available in the future). At this
moment, our procedure is similar to that in \citet{hs2022}. We consider the frequency $\nu$ as an
auxiliary variable only, and use Eq.~(\ref{prec}) to connect it with parameters of more interest,
such as the stellar masses or mutual inclination $J$ of the inner and outer orbital planes. This
does not mean that the new parameters would have a fully correlated solution. This is because (i)
$J$, for instance, depends on ${\bf {\hat l}}_1$ and ${\bf {\hat l}}_2$ on a more fundamental
level (recall $\cos J = {\bf {\hat l}}_1\cdot {\bf {\hat l}}_2$), and (ii) constraints following
from the spectroscopic observations help to decorrelate the stellar masses. So in principle, all
quantities on the right hand side of the formula (\ref{prec}) may be adjustable in the process of
data fitting. However, for some of them we assume that their pre-constrained values from the
analysis in Sec.~\ref{obs} are precise to such a level that the data describing secular evolution
of the system would not result in improvement (and in the same time, their uncertainty would not
significantly affect solution of the parameters which are fitted at this stage). For the sake of
simplicity, we thus adopt their best-fit values and consider them constant. This is the case of
the mean orbital periods $P_1$ and $P_2$ of the inner and outer orbits, and the eccentricity $e_2$
of the outer orbit. In fact, the latter appears in $\nu$ and $L_2$ only via the $\eta_2$ factor,
and therefore depends on $e_2^2$ rather than $e_2$. We are then left with the stellar masses, of
which $m_2$ is perhaps the most interesting. This is because the analysis of the spectroscopic
data in Sec.~\ref{spectra} helps to directly constrain both $m_{1a}$ and $m_{1b}$, since $i_1$ is
well known at the epoch of the spectroscopic observations from the conjoined photometric
observations of the eclipsing system. A similar determination of $m_2$ from spectroscopy would
need $i_2$ to be known, but this is available from the long-term evolution model only (and not
directly from observations). Given the still limited quality of available data, we thus resolve to
provide a preliminary solution for the V907~Sco system at this moment. We fixed $m_{1a}$ and
$m_{1b}$ values determined in Sec.~\ref{phot}, and we adopt $m_2$ as the fourth solved for
parameter.
\begin{figure*}[ht!]
 \epsscale{1.}
 \plotone{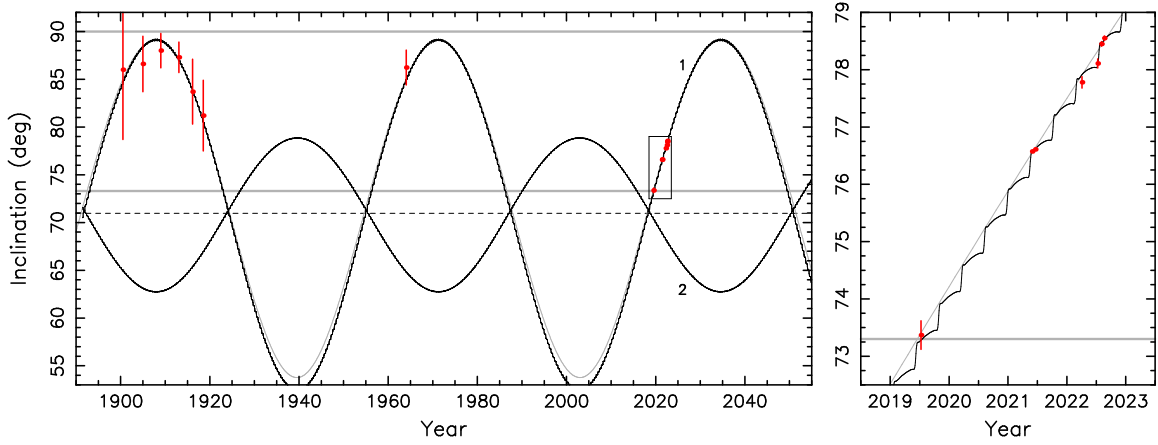}
 \caption{Inclination $i_1$ (label~1) and $i_2$ (label~2) of the inner- (binary) and the outer-component
  orbits of V907~Sco over the past century and extending towards the next decade (the abscissa
  in calendar years). Red symbols are data with uncertainties determined in Sec.~\ref{obs} (Table~\ref{inclination}).
  Right panel provides a zoom on the modern, most accurate data (depicted using the rectangle
  on the right panels) and shows only the $i_1$ values. The best-fitting model from
  Figure~\ref{fig1} (blue star) having $m_2=1.062$~M$_\odot$, $P_\nu=63.5$~yr and $J=26.2^\circ$. The
  grey horizontal line at inclination $73.3^\circ$ delimit the zone of $i_1$ values for which the
  inner binary is eclipsing, the one at $90^\circ$ shows just the configuration maximizing eclipses
  depth. The black dashed horizontal line at inclination $71^\circ$ shows the inclination of the
  total angular momentum of the system, about which the inner and outer orbital planes precess.
  The grey line is the approximate best-fit solution from the secular
  model from Sec.~\ref{theory1} only. The full solution (solid line and Sec.~\ref{theory2}) contains
  also the long-periodic signal with periodicity $P_2$, required to match the data from 2022.}
 \label{fig2}
\end{figure*}

\subsection{Numerical model for the long-period part} \label{theory2}
We used the secular model from Sec.~\ref{theory1} to fit the photometric observations
(Sec.~\ref{res_sec} and Fig.~\ref{fig2} in particular) and reached the following conclusions: (i)
the model helps us to match the mean trend of the data adequately, and allows us to characterize a
large portion of the parameter space which cannot lead to satisfactory results (thus pre-constrain
the parameter space), but (ii) formally the best fit is statistically not acceptable (the $\chi^2$
from Eq.~(\ref{chi2}) is simply too large). While looking into details, we noticed the reason for
the situation: on top of the secular terms, there are also long-period perturbations (periodicity
$P_2$ and its multiples) that contribute in a non-negligible way to the long-term evolution of
$i_1$ and $i_2$ values. Since the post-2019 photometric data are rather accurate, it is important
to include these terms into the analysis. At the lowest-orders in the outer eccentricity $e_2$
they can be expressed analytically, e.g., \citet{s1975} \citep[but see already][]{b1936}, and
represent a signal with periodicity $P_2/2$. For larger $e_2$ values, which is the case of
V907~Sco, an accurate analytic formulation is not available to us. For that reason, we decided to
resolve the situation by using a fully numerical model, and profit from the a priori parameter
constraints from the secular model described above. Interestingly, we find that at high $e_2$
regime the leading periodicity of the long-term perturbation of both $i_1$ and $i_2$ becomes $P_2$
(Fig.~\ref{fig2}).\footnote{
 At the resubmission stage we noted that
 \citet{betal2011} and \citet{betal2015}, Appendix D in particular, provide both justification for the change in
 the period of the dominant perturbing term in $i_1$ when $e_2$ is large, but also an analytic description of the corresponding long-period effect. Our numerical approach reproduces the results accurately.}
The set-up of the numerical model uses a point mass configuration and Jacobi coordinates
\citep[e.g.,][Eqs.~(1)]{s1982}. The near-coplanarity of the inner- and outer-orbits in the
V907~Sco system also does not require us to include the gravitational interaction of the stars in
the eclipsing system as extended bodies (beyond the point-mass level) that would be otherwise
needed for long-term stability if $J$ is large. None of the observable quantities depends on these
three-body effects. Because the long period perturbations depend on the mean longitude in orbit of
the outer system at the reference epoch $T_0$, we must add it to the set of adjustable parameters.
Taking the argument of pericenter $\omega_2$ from the analysis of radial velocities
(Sec.~\ref{spectra}), we let the mean anomaly $\ell_2$ be the fifth solved-for parameter. There is
no a priori constraint on $\ell_2$, and its value is allowed to sample the full interval of
$360^\circ$.

\begin{figure*}[ht!]
 \epsscale{0.9}
 \plotone{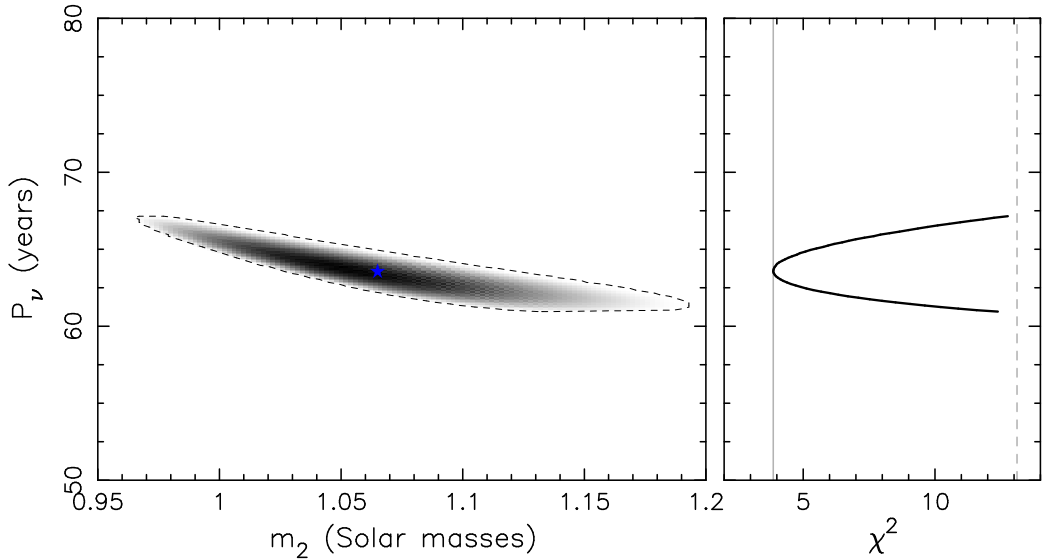}
 \caption{Left: distribution of solutions for which $\chi^2\leq \chi^2_\star\simeq
  13.13$ projected
  onto the plane of mass $m_2$ (abscissa) and precession period $P_\nu$ (ordinate). Solution
  density is indicated by greyscale, white for no possible
  solution, black for the largest number of solutions. The dashed lines delimit the zone
  of acceptable solutions. The blue star at $m_2=1.062$~M$_\odot$ and $P_\nu=63.5$~yr shows
  location of the best-fitting solution with $\chi^2_{\rm min}=3.89$. Right: the
  minimum $\chi^2$ (abscissa) for a given $P_\nu$ value (ordinate). The dashed
  vertical line is the limit for accepted solutions $\chi^2=\chi^2_\star$, the
  solid vertical line is the best achieved value $\chi^2_{\rm  min}\simeq 3.89$.}
 \label{fig1}
\end{figure*}

\subsection{Dataset for long-term analysis}
\begin{figure}[t!]
 \plotone{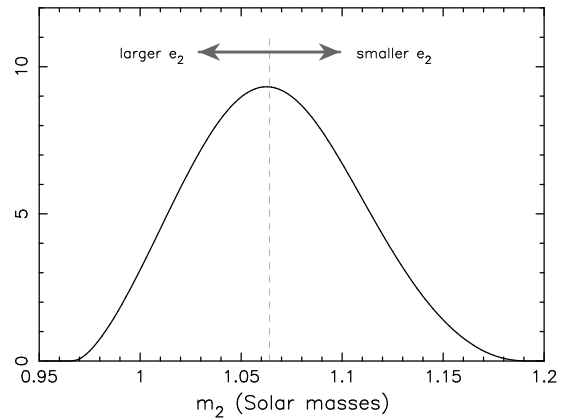}
 \caption{Probability density distribution of solutions for mass $m_{2}$ of the third component
  in the V907~Sco system. The available dataset of photometric observations was used. The non-zero
  values correspond to the projection of five-dimensional parameter-space zone ${\cal V}$ containing
  all admissible solutions within 90\% confidence limit. The vertical dashed line is the median value of
  the distribution. Should the eccentricity $e_2$ of the outer orbit be smaller/larger than the
  nominal value $0.467$ used here, the distribution would shift towards larger/smaller values
  as shown by the arrows.}
 \label{fig3}
\end{figure}

\begin{figure*}[t!]
 \plottwo{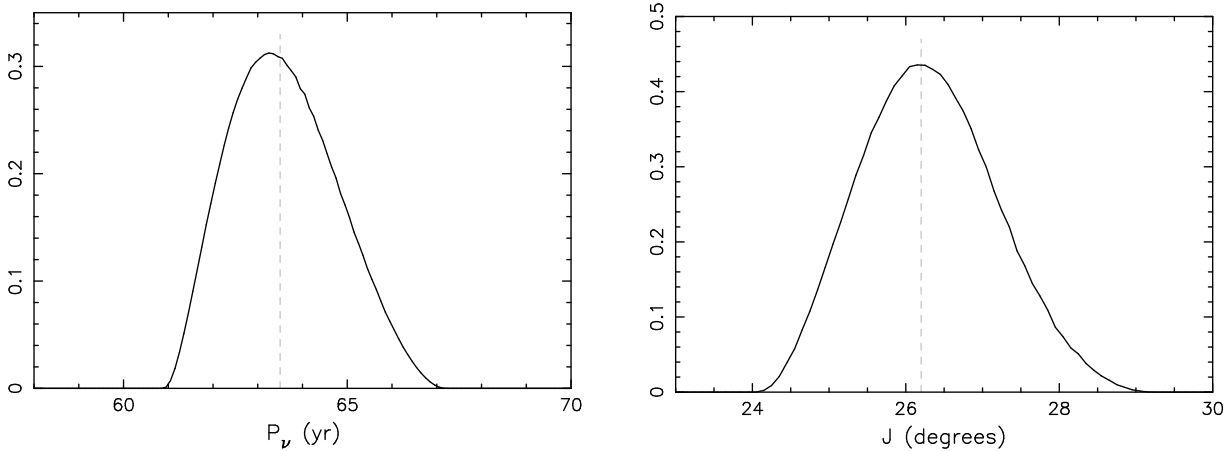}{f5.eps}
 \caption{Probability density distribution of solutions for precession period $P_\nu$ and mutual
  inclination $J$ of inner and outer orbits of the V907~Sco system. The available dataset of
  photometric observations was used. The non-zero values correspond to the
  projection of five-dimensional parameter-space zone ${\cal V}$ containing all admissible solutions
  within 90\% confidence limit. The vertical solid line is the median value of the distribution.}
 \label{fig4}
\end{figure*}

Data suitable for long-term analysis of the evolution of V907~Sco consist of results from our
analysis of the photometric and spectroscopic observations at individual and separated intervals
of time presented in Sec.~\ref{obs}. Each of them provided at a certain epoch $t_j$, usually at
the center of the relevant time interval, a value of a specific variable $v_j$ together with its
uncertainty $\Delta v_j$. The triples $(t_j,v_j,\Delta v_j)$, with $j=1,\ldots,N_{\rm data}$, are
the basis of our work in this section. The model described in Secs.~\ref{theory1} and
\ref{theory2} allows us to compute the values ${\hat v}_j$ of these variables at the same epochs
$t_j$, and we can thus use the standard $\chi^2$ metric defined by
\begin{equation}
 \chi^2 = \sum_{j=1}^{N_{\rm data}} \left(\frac{v_j-{\hat v}_j}{\Delta v_j}\right)^2\; ,
  \label{chi2}
\end{equation}
as a measure of the model quality. As expected, ${\hat v}_j={\hat v}_j({\bf p})$, where ${\bf
p}=({\hat {\i}}_1,{\hat {\i}}_2,{\hat \Omega}_2,m_2,\ell_2)$ constitute the five-dimensional
parametric space. Our procedure obviously seeks minimization of $\chi^2({\bf p})$, reaching a
minimum value $\chi^2_{\rm min}$ for a certain, best-fit set of parameters ${\bf p}_{\rm min}$.
Making sure that the fit is statistically acceptable \citep[to that end we use the $Q$-function
test, see][Chap.~15.1]{nr2007}, we seek confidence limits on estimated model parameters ${\bf p}$.
This is achieved by selecting a certain volume ${\cal V}$ in the parameter space characterized by
$\chi^2\leq \chi^2_\star=\chi^2_{\rm min}+\Delta\chi^2$. For a five-dimensional parameter space,
and a $90$\% confidence level limit, we have to choose $\Delta \chi^2=9.24$ \citep[see,
e.g.,][Chap.~15.6]{nr2007}. Projection of ${\cal V}$ onto lower-dimensional sections of the
parameter space allows us to re-map the statistical analysis onto other variables of interest,
which do not constitute the basic set of dimensions by ${\bf p}$ (for instance, we may
characterize this way acceptable values of the precession period $P_\nu=2\pi/\nu$, as shown in
Figs.~\ref{fig1} and \ref{fig4}, or the mutual angle $J$ of the orbital planes, as shown in
Fig.~\ref{fig4}). Even simpler is just projecting ${\cal V}$ onto the individual axes defined by
each of the parameters ${\bf p}$ (for instance to infer limits in the companion mass $m_2$,
Figs.~\ref{fig1} and \ref{fig3}).

Having outlined our procedure, we now return to the choice of observables $v_j$. Their sector
based on photometric observations consists of data summarized in Table~\ref{inclination}, namely
the inferred inclination values $i_1$ of the inner binary orbit from mutual eclipses. This
provides 14 datapoints. Next, the spectroscopic observations taken at CTIO in 2022 provide further
constrains. First, the analysis of the radial velocity semi-amplitudes of the stars in the
eclipsing binary (Fig.~\ref{RV_V907Sco}), and the known inclination $i_1$ at their epoch, provides
values of masses $m_{1a}$ and $m_{1b}$. We consider them fixed. The analysis of the spectroscopic
observations further provides radial velocity data for the outer orbit with period $P_2$, namely
how it affects the systemic velocity of the barycenter of the eclipsing binary
(Fig.~\ref{RV3plot}). This provides a constraint on the argument of pericenter $\omega_2$ and the
eccentricity $e_2$ that we also consider fixed in our analysis (we only tested the influence of
making $e_2$ smaller or larger than nominal). Being most interested in constraining the mass $m_2$
of the third, unseen star, we observe the semi-amplitude of the radial velocity curve in
Fig.~\ref{RV3plot}, from which we can derive the projected semimajor axis $a_{2,{\rm
proj}}=a_2\,(m_2/M_2)\,\sin i_2$ (the semimajor axis of the outer orbit is given by the Kepler's
third law $n_2^2 a_2^3 = G M_2$). We obtain, $a_{2, {\rm proj, obs}}(t_s) = 34.14$~R$_\odot$ with
an uncertainty $\Delta a_{2,{\rm proj, obs}} = 2.04$~R$_\odot$ at a reference HJD epoch
$t_s=59760$. This constitutes our additional data to be fit by our model. Finally, we chose the
reference epoch $T_0=59383.9$ heliocentric Julian date, namely the mid epoch of the TESS
sector~39b observations (Tab.~\ref{inclination}).

\subsection{Results} \label{res_sec}
We ran two sets of simulations: (i) first using only the photometric data, and (ii) second using
both photometric data and spectroscopic constraints on the projected semimajor axis of the outer
orbit. Our intention was to investigate how good the solution was using the photometric data
alone, and how it potentially improves if constraints from spectroscopy are added.

Figure~\ref{fig2} shows the best-fitting solution of the inclination $i_1$ data, including also
coupled variations of the outer orbit inclination $i_2$. The inclination of the conserved total
angular momentum (i.e., the invariable plane) is $\simeq 71^\circ$, indicating the solution
belongs to the class 2 discussed by \citet{letal1999} in Sec.~4 of their analysis (which they
rightly favoured). The archival, pre-2000 data are formally inaccurate, nevertheless they help to
significantly constrain the solution. The modern data, especially those after 2021, are very
accurate and help constrain the solution even more. The important advantage of V907~Sco over the
HS~Hya case \citep{hs2022} is the fact that the data cover a time interval much longer than the
precession period $P_\nu=2\pi/\nu$. This makes the free-parameter values much better constrained:
compare Fig.~\ref{fig1} here with Figs.~3 and 4 in \citet{hs2022}. Figure~\ref{fig3} presents the
probability density distribution of the third mass $m_2$. We obtain
$m_2=1.06^{+0.11}_{-0.10}$~M$_\odot$, representing a $90$\% confidence level interval. We briefly
tested the sensitivity of the solution to changing the value of the eccentricity $e_2$ of the
outer orbit (which has been fixed in the illustrated solution to its nominal value $0.467$). We
found that slightly smaller or larger $e_2$ values would shift the $m_2$ solution to slightly
larger or smaller values. This is to be expected. For instance, smaller $e_2$ would place the
outer star more distant to the eclipsing binary. In order to produce the observed variations of
$i_1$ values, one would thus need larger mass $m_2$ (and vice versa). Figure~\ref{fig4} presents
the probability density distribution of the precession period $P_\nu$, reflected in periodicity of
variations of $i_1$ and $i_2$ in Fig.~\ref{fig2}, and the mutual angle $J$ of the inner and outer
orbits orbital planes. The 90\% confidence values are $P_\nu=63.5^{+3.3}_{-2.6}$~yr and
$J=26.2^{+2.6}_{-2.2}$ degrees. The value of $J$ is obviously the sum of the semiamplitudes of
$i_1$ and $i_2$ variations seen in Fig.~\ref{fig2}, while the ratio of their sine values is
$\gamma\simeq 0.47$ from Eq.~(\ref{gama}). The $J$ value is just large enough to produce
interesting orbital effects, but smaller than the critical value of $\simeq 40^\circ$ for the
onset of Kozai-Lidov effects \citep[e.g.,][]{s1982,fl2010}. The use of a simple point-mass model
in our simulations is therefore well justified. We also note that the argument of pericenter of
the outer orbit exhibits a drift of $\simeq 1.35$~degrees per year (this is not an observationally
detected value yet, but it follows from our numerical simulation; see also Eq.~(8) in
\citet{hs2022}). This large value stems from compactness of the system ($P_2/P_1$ not too large)
and large mass of the binary. This effect needs to be taken into account, apart from the changing
inclination values $i_1$ and $i_2$, if we were to interpret the radial velocity measurements
reported in \citet{letal1999} (Table~3; see also Sec.~\ref{concl}).

Next, we added to the dataset the projected semiamplitude $a_{2,{\rm proj}}$ constraint from the
spectroscopic data in 2022. We found that this single data point did not improve the solution
meaningfully at this moment. This is because (i) the photometric data already constrain the model
very well, and (ii) the spectroscopic information is not very accurate yet, principally because
the observations missed to detect precise moment of the pericenter passage of the outer orbit
(Fig.~\ref{RV3plot}).

\section{Discussion and conclusions} \label{concl}
The recent onset of eclipses of the binary component of the V907~Sco system recorded in the TESS
data opened a new opportunity to study this interesting triple. This work is only an initial
attempt in this direction, but it nevertheless reveals important corrections to prior results.
They are as follows:
\begin{itemize}
\item masses of stars in the eclipsing binary, $2.74\pm  0.02$~M$_\odot$ and $2.56\pm
0.02$~M$_\odot$,
 are about $8$\% larger than previously thought; this is because \citet{letal1999} assumed
 in their work an inaccurate value of the inclination $i_1$ of the binary orbital plane at the
 effective epoch of their spectroscopic observations;
\item our single-season spectroscopic data and a more precise modeling work allowed us to
constrain parameters of the orbit of the
 third star about the center of mass of the binary; unlike previously thought, we
 find it has a significant eccentricity of $0.47\pm 0.02$ and a substantially longer period of
 $142.01\pm 0.05$~days;
\item secular and long-period perturbations in $i_1$ produced by the third component in the
 system allow us to constrain its mass of $1.06^{+0.11}_{-0.10}$~M$_\odot$ and determine mutual
 inclination of the two orbital planes of $J=26.2^{+2.6}_{-2.2}$ degrees.
\end{itemize}
The sum of all known parameters, both orbital and physical, is given in Tables~\ref{LCaRVparam}
and \ref{RVpar3}. In principle, one may use those data to estimate, at least at zero order, the
age of stars in the eclipsing binary. However, we find that the result may sensitively depend on
their yet unknown metalicity. In any case, the values range in between $\simeq 300$ and $\simeq
500$~Myr. This would mean that V907~Sco is a rather young stellar system. More detailed analysis
of this issue, including consistency with the estimate of a circularization timescale for the
inner binary orbit, is left for future efforts.

\begin{figure}[t!]
 \plotone{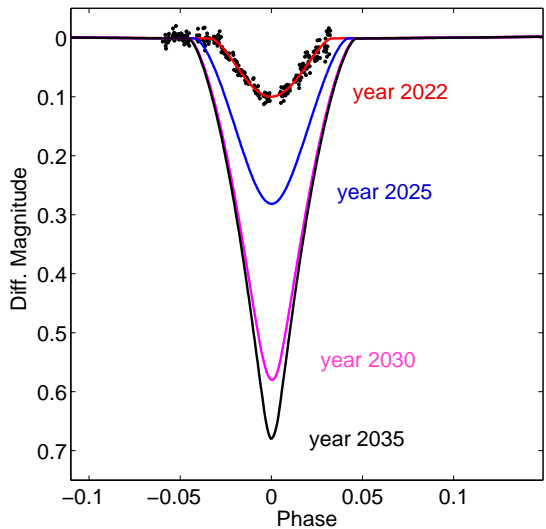}
 \caption{Predicted increase of the primary eclipse depths for the V907~Sco inner binary
  in the next decade. The effect is due to increasing inclination $i_1$ of the orbital
  plane (Fig.~\ref{fig2}). The red line is adjusted to the actual measurements in
  2022 reported in Sec.~\ref{obs}. The abscissa gives the phase of the $P_1\simeq 3.776$~day
  cycle, the ordinate is magnitude.}
 \label{fig_depth}
\end{figure}

The archival and incidental photometry data of V907~Sco proved to be valuable for description of
the long-term changes in its architecture. However, their accuracy is obviously low. Modern
instrumentation, even of a small aperture, may provide much better results if the star is
purposely targeted. Additionally, the situation of V907~Sco is fortuitous since the inclination
$i_1$ will be increasing to a nearly $90^\circ$ value for the next decade and the system will be
eclipsing until the year $\sim2049$ (Fig.~\ref{fig2}). Therefore the upcoming years may offer a
chance to significantly improve our understanding of this unique system. As an example,
Fig.~\ref{fig_depth} shows prediction of the increasing depth of primary eclipses during the next
decade, a trend reflecting the increase of the inclination $i_1$ from Fig.~\ref{fig2}. This
progression holds the promise for $i_1$ to be determined more precisely in the coming years and
for the results of this paper to be improved upon rapidly.

The unfortunate miss of the outer orbit radial velocity minimum (Fig.~\ref{RV3plot}) in our work
strongly motivates further efforts to observe the V907~Sco system spectroscopically. An intense
campaign around the epoch of this minimum (the nearest chance in May~2023) would be especially
valuable. These data will help to pin down values of both outer orbit period $P_2$ and its
eccentricity $e_2$. Additionally, if the projected semimajor axis of the outer orbit $a_{2,{\rm
proj}}$ is better constrained from these future data, the analysis developed in this paper may
readily provide much better constraints of the mass $m_2$.

We also give a brief look at expected, but not detected yet, eclipse time variations (ETVs) in the
inner binary of V907~Sco. As the eclipse depth will continually increase in the future
(Fig.~\ref{fig_depth}), determination of the mid-eclipse epochs will improve, thus permitting the
discovery of variations with respect to the mean period $P_1$. If accurate enough, proper
interpretation of ETVs may help to further constrain the physical parameters of the system. There
are two kinds of ETVs, and they are distinguished by their physical origins. First, the light-time
(Roemer) effect is due to the finite speed of light and variations of radial distance (projected
along the line of sight) to the eclipsed component in the binary due to its motion about the
barycenter of the whole triple system. Second, the dynamical (physical) effect has to do with
gravitational perturbation of the third star of the mean motion of stars in the eclipsing binary.
Analytical description of the former is rather straightforward, while the same for the latter
requires a more involved technique of the perturbation theory to be used. Over the past decade or
so, a thorough analysis has been developed by Borkovits and collaborators \citep[e.g.,][and
references therein]{bor2022}. Here we specifically used two components of the dynamical ETV signal
given in Eqs.~(8) and (9) in \citep{retal2013}, and we verified that additional parts are very
small. To make sure that this formulation is sufficient, we also determined the dynamical part of
the ETVs using direct numerical integration. This is basically the same simulation as shown in
Fig.~\ref{fig2}, but here we paid special attention to the epochs of eclipses of the two stars in
the inner binary.

\begin{figure}[t!]
 \plotone{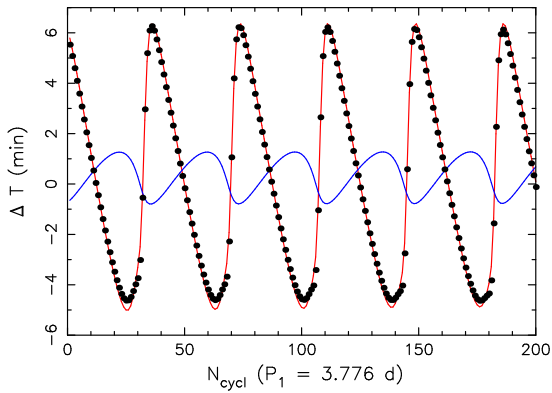}
 \caption{Predicted value of ETVs ($\Delta T$) for the primary component in the V907~Sco. The
   abscissa is the cycle count of the eclipsing binary (period $P_1=3.776$~day; the
   whole timespan shown is therefore little more than two years),
  the ordinate is in minutes. The red line is the dynamical (physical) delay due
  to the third-star perturbations of the mean motion of the binary, the blue line
  is the light-time (Roemer) effect due to motion of the binary center of mass
  about the barycenter of the whole system. Both are expressed analytically
  \citep[e.g.,][]{retal2013}. Symbols show the dynamical delay determined numerically
  for sake of comparison. The perihelion of the outer orbit motion occurs where the
  gradient of the dynamical effect change is the largest.}
 \label{fig6}
\end{figure}

Results are shown in Fig.~\ref{fig6}. We note the dynamical effect dominates by a factor of $\sim
5$ over the light-time effect. This is because the system is compact ($P_2/P_1$ small enough) and
the mass of the third star not too small ($m_2/M_2$ large enough). The full amplitude of the
dynamical effect, more than $10$~minutes, appears large enough to be safely detectable. Especially
in the near future when the eclipses will be deep enough, and also the ground-based data would
provide adequately precise times of eclipses (which is not the case nowadays, and the number of
TESS data is still too limited).

At this moment, we did not attempt to further constrain our solution for the orbital architecture
of V907~Sco by using radial velocities published by \citet{letal1999}. In principle, this is
possible by simple propagation of our numerical model back to the epochs of the spectroscopic
observations from Table~3 in \citet{letal1999}, though one would need to assume some uncertainty
of the reported radial velocity values. We plan to return to this issue after the system is better
constrained by well controlled and accurate photometric and spectroscopic observations in the next
year. In particular, V907~Sco will be in the field of view of TESS again in Sector~66 of its
operations. This will provide very accurate photometric data. Second, we now have a fairly good
constraint when the system will pass through the pericenter of the outer orbit. We can thus make
efforts to give a high priority to observation during that time period. Such targeted
spectroscopic observations will help to improve parameters of the outer orbit, and consequently to
constrain the mass $m_2$ of the third star in the system.

Assuming our solution for $m_2\simeq 1.06$~M$_\odot$ is correct, one may question the role of the
radiative flux of the third component in the V907~Sco system in our analysis of the lightcurves
during eclipses of the inner binary (the so called third light problem). Given the masses $m_{1a}$
and $m_{1b}$ in the range $2.54$ to $2.74$~M$_\odot$, we can estimate the third light contribution
at the level $5$~\%. This is in a good accordance with the TESS lightcurve analysis in
Sec.~\ref{obs}, which indicated the third light should be around this value. In the same way, the
small $\simeq 0.008$ eccentricity of the inner binary is close to the forced value by the
companion star of approximately solar mass and the orbital parameters listed in Table~\ref{RVpar3}
\citep[see, e.g.,][end references therein]{ger2009}.

To conclude with a speculative comment, we note that there is a field star UCAC2 17118369  =
2MASS J17565485-3444584 separated from V907~Sco by $\simeq 9^{\prime\prime}$ \citep[possibly the
object shown by][in their Fig.~1]{letal1999}. Interestingly, the WDS catalogue records the two
stars to be a proper-motion pair \citep{2001AJ....122.3466M}. More importantly though, the Gaia
DR3 provides (i) the same distances to both objects within their uncertainty, and (ii) confirms
very similar values of their proper motion \citep[though in this case slightly outside their
uncertainty overlap,][]{2022arXiv220800211G}. A detailed look into the putative gravitational
bound between the two stars is beyond the scope of the present paper. Here we only mention that
(i) their $\simeq 3.5$ magnitude difference contributes insignificantly, if any, to our TESS
photometry analysis in Sec.~\ref{obs}, and (ii) the large distance of the companion cannot have
influence on our analysis of the V907~Sco triple dynamics.

\begin{acknowledgments}
 We would like to thank an anonymous referee for his/her helpful and critical suggestions improving the level of the manuscript.
 We thank T. Borkovits for drawing our attention to the TESS 2019 observations of the V907~Sco eclipses
onset, P. Harmanec for useful discussions and checks, and the referee, whose suggestions helped to
improve the final version of this paper. The work of DV was partially supported by the Czech
Science Foundation (grant~21-11058S). BB was supported by National Science Foundation grant
AST-1812874.
 The DASCH project at Harvard is grateful for partial support from NSF grants AST-0407380, AST-0909073,
and AST-1313370.
 Funding for APPLAUSE has been provided by DFG (German Research Foundation, Grant), Leibniz Institute
for Astrophysics Potsdam (AIP), Dr. Remeis Sternwarte Bamberg (University Nuernberg/Erlangen), the
Hamburger Sternwarte (University of Hamburg) and Tartu Observatory. Plate material also has been
made available from Th\"uringer Landessternwarte Tautenburg, and from the archives of the Vatican
Observatory.
 We would also like to thank the Pierre Auger Collaboration for the use of its facilities. The operation
of the robotic telescope FRAM is supported by the grant of the Ministry of Education of the Czech
Republic LM2018102. The data calibration and analysis related to the FRAM telescope is supported
by the Ministry of Education of the Czech Republic MSMT-CR LTT18004, MSMT/EU funds
CZ.02.1.01/0.0/0.0/16\_013/0001402 and CZ.02.1.01/0.0/0.0/18\_046/0016010.
 This work has made use of data from the European Space Agency (ESA) mission Gaia
(\url{https://www.cosmos.esa.int/gaia}), processed by the Gaia Data Processing and Analysis
Consortium (DPAC, \url{https://www.cosmos.esa.int/web/gaia/dpac/consortium}). Funding for the DPAC
has been provided by national institutions, in particular the institutions participating in the
Gaia Multilateral Agreement.
\end{acknowledgments}

\bibliography{stars}{}
\bibliographystyle{aasjournal}

\appendix

\section{New radial velocity measurements for eclipsing binary in V907~Sco}

\startlongtable
\begin{deluxetable*}{ccccc}
\tablenum{A1} \tablecaption{Radial velocities, and their uncertainty, of the stars
 in the inner binary of V907~Sco measured by {\sc reSPEFO 2} package.
 Data obtained at CTIO observatory using CHIRON echelle spectrograph
 in the 2022 season (March to October).
 \label{tableRVdata}}
\tablewidth{0pt} \tablehead{
 \colhead{Time} & \colhead{RV$_{1}$} & \colhead{$\sigma_{1}$} & \colhead{RV$_{2}$} & \colhead{$\sigma_{2}$} \\
 \colhead{(HJD-2400000)} & \colhead{(km s$^{-1}$)} &
 \colhead{(km s$^{-1}$)} & \colhead{(km s$^{-1}$)} &
 \colhead{(km s$^{-1}$)}
}
\startdata
  59667.8470 &  +51.6 &   0.9 &    -80.5 &   0.7 \\
  59667.8886 &  +44.8 &   0.6 &    -73.5 &   0.5 \\
  59668.8605 & -123.4 &   0.7 &    +84.8 &   0.7 \\
  59668.8936 & -126.1 &   0.5 &    +87.7 &   0.6 \\
  59670.8707 &  +99.7 &   0.7 &   -121.5 &   0.4 \\
  59671.8247 &  +17.0 &   0.4 &    -43.4 &   0.3 \\
  59672.8766 & -134.1 &   0.5 &    +98.1 &   0.5 \\
  59673.8782 &  -13.3 &   0.7 &     --   &   --  \\
  59674.8637 & +106.5 &   0.8 &   -124.4 &   0.6 \\
  59675.8488 &  -31.2 &   0.6 &     +5.3 &   0.7 \\
  59683.7508 &  -92.1 &   0.7 &    +65.3 &   0.6 \\
  59683.9130 & -113.9 &   0.6 &    +85.3 &   0.5 \\
  59684.7279 &  -93.7 &   0.7 &    +65.7 &   0.7 \\
  59684.8764 &  -70.2 &   0.8 &    +43.7 &   0.5 \\
  59685.7483 &  +84.0 &   0.9 &    -98.6 &   1.0 \\
  59685.9033 &  +99.2 &   0.5 &   -112.9 &   0.5 \\
  59686.7077 &  +60.7 &   0.6 &    -77.0 &   0.7 \\
  59686.9322 &  +19.7 &   0.5 &    -39.7 &   0.8 \\
  59687.7161 & -115.9 &   0.6 &    +88.2 &   0.7 \\
  59687.9335 & -129.7 &   0.7 &   +101.2 &   0.6 \\
  59688.7270 &  -56.3 &   1.1 &    +32.2 &   1.0 \\
  59689.7208 & +101.7 &   0.7 &   -115.2 &   0.6 \\
  59689.8543 & +108.1 &   0.7 &   -121.1 &   0.7 \\
  59690.7445 &  +12.9 &   0.9 &    -31.8 &   0.8 \\
  59690.8288 &   -2.4 &   1.3 &    -15.2 &   1.5 \\
  59690.8773 &  -14.9 &   1.5 &     -8.8 &   1.7 \\
  59690.9364 &  -24.5 &   1.5 &     +5.6 &   1.5 \\
  59691.6965 & -128.3 &   0.7 &    100.9 &   1.4 \\
  59691.8703 & -128.8 &   0.9 &    100.4 &   1.3 \\
  59691.9332 & -127.6 &   1.4 &     99.2 &   1.2 \\
  59692.7019 &  -17.3 &   0.5 &     -3.6 &   0.5 \\
  59692.8727 &  +14.9 &   0.6 &    -33.8 &   0.6 \\
  59693.7070 &  110.0 &   0.7 &   -121.2 &   0.8 \\
  59693.8903 &  103.7 &   0.8 &   -115.9 &   0.7 \\
  59724.7336 &  +12.2 &   0.7 &    -31.8 &   0.7 \\
  59724.8911 &  -21.1 &   2.2 &     -1.7 &   1.2 \\
  59726.6085 &  -33.7 &   1.0 &    +10.1 &   0.8 \\
  59726.8641 &  +15.9 &   1.0 &    -36.0 &   1.0 \\
  59727.6148 & +107.9 &   0.8 &   -121.6 &   0.7 \\
  59727.8497 & +103.3 &   0.9 &   -118.7 &   0.7 \\
  59738.6036 &  +80.6 &   0.8 &   -100.1 &   0.7 \\
  59738.7404 &  +95.2 &   0.8 &   -113.1 &   0.6 \\
  59741.6162 &  -54.3 &   0.8 &    +25.0 &   0.7 \\
  59741.8417 &  -13.2 &   0.5 &    --    &   --  \\
  59742.6063 & +100.3 &   0.8 &   -120.3 &   0.8 \\
  59742.8642 & +105.6 &   0.7 &   -124.4 &   0.7 \\
  59794.5385 &  -65.3 &   0.6 &     -5.4 &   0.6 \\
  59794.7500 &  -23.0 &   1.3 &    -43.9 &   1.0 \\
  59795.5210 &   83.6 &   0.7 &   -141.7 &   0.6 \\
  59795.7135 &   85.8 &   1.1 &   -144.0 &   0.9 \\
  59796.5105 &  -14.7 &   0.5 &    -47.7 &   0.5 \\
  59796.6964 &  -52.4 &   0.6 &    -12.5 &   0.4 \\
  59797.5036 & -151.1 &   0.5 &     82.3 &   0.5 \\
  59797.6887 & -146.6 &   0.5 &     78.6 &   0.6 \\
  59799.5262 &   90.7 &   0.5 &   -138.3 &   0.5 \\
  59799.7060 &   79.3 &   1.1 &   -128.3 &   1.0 \\
  59800.5431 &  -61.1 &   0.5 &      6.2 &   0.5 \\
  59800.7231 &  -93.9 &   0.7 &     36.7 &   0.9 \\
  59801.5195 & -138.0 &   0.5 &     79.8 &   0.5 \\
  59801.7196 & -118.1 &   0.5 &     61.0 &   0.6 \\
  59802.5259 &   27.6 &   0.5 &    -71.3 &   0.4 \\
  59802.7273 &   60.9 &   0.9 &   -101.3 &   0.6 \\
  59840.5154 &   77.6 &   0.7 &    -91.7 &   0.5 \\
  59848.5099 &  110.2 &   0.6 &   -121.5 &   0.4 \\
  59859.5077 &   90.3 &   0.7 &   -104.0 &   0.5 \\
  59866.4861 &   -3.1 &   0.8 &    -17.7 &   0.7 \\
\enddata
\end{deluxetable*}

\end{document}